\documentclass[prd,twocolumn,showpacs,showkeys,preprintnumbers]{revtex4}
\usepackage{amsmath,epsfig}
\newcommand{\Ds}{\text{$/$\hspace{-0.37cm} $\nabla$}}
\newcommand{\Z}{{\bf Z}}
\newcommand{\Tr}{{\rm Tr}}

\begin{document}
\title{Unification in Intersecting Brane Models}
\author{Kang-Sin Choi}
\email{kschoi@th.physik.uni-bonn.de}
\affiliation{Physikalishes Institut, Universit\"at Bonn, Nussallee
 12, D-53115 Bonn, Germany}
\affiliation{Center for Theoretical Physics, Seoul National
  University, Seoul 151-747, Korea}
\preprint{SNUTP 06-002}

\begin{abstract}
We propose a unification scenario for supersymmetric intersecting
brane models. The quarks and leptons are embedded into adjoint
representations of $SO(32)$, which are obtained by type I string and
broken by compactification on orbifolds. Its single unified gauge
coupling can give rise to different gauge couplings below the unification
scale, due to effects of magnetic fluxes.
The crucial mechanism is brane recombination preserving supersymmetry.
\end{abstract}

\maketitle

Although the open string description of the gauge theory in type I/II brane
stack models provides a promising explanation of the origin of Minimal
Supersymmetric Standard Model (MSSM),
there have been lack of understanding on unification, contrary to
other compactification schemes. If various vacua
are to be understood as spontaneously broken phases, above the
compactification scale, of a string theory, there should be an
unification of gauge group and gauge coupling. In
this letter we show that indeed gauge groups and couplings unify
described by brane recombination preserving supersymmetry.

\section{Embedding into SO(32) adjoints}

One may note that, for example, the $X,Y$ gauge bosons in conventional
Georgi--Glashow $SU(5)$ unification \cite{GG} have the correct
 colors and weak isospins $\bf (3,2)$,
 as (s)quarks. Of course
they do not have the desired $U(1)$ hypercharges, but
provided that they are also charged under additional $U(1)$ symmetries,
after diagonal symmetry breaking they may have the correct quantum
numbers. We first investigate this possibility: embedding all the
MSSM fields into the {\em $SO(32)$ adjoints}, which are the only
representations predicted by type I string theory \footnote{This is also
  along the line of gauge--matter--Higgs unification ideas \cite{GLMN}.}.
Clearly, however, we require much bigger group than
Georgi--Glashow $SU(5)$.
The ingredients are extra dimensions and magnetic fluxes, which is
equivalent to intersecting branes at angles.
Masslessness and chirality come from zero modes of Dirac operator.

A supporting evidence of this picture is suggested in the dual
{\it M/F}-theory compactification \cite{AW}. The
intersection of D6 branes is purely geometrically described as
unwinding of $U(M+N)$ singularity to $U(M)$
and $U(N)$ and the bifundamental fermions come from the
branching of the adjoint
\begin{equation*}
 \bf (M+N)^2 \to (M^2,1) +(1,N^2) + (M,N)
\end{equation*}
where we should count only one bifundamental because the other is $CPT$
conjugate. In the intersecting brane picture,
a string stretched from $M$ brane stack to $N$ brane stack
corresponds to {\em chiral} bifundamental representation
$\bf (M,N)$, and the other string having the opposite
orientation is $CPT$ conjugate due to the opposite
GSO projection \cite{BDL}.

Consider a toy model.
The gauge group $H=U(3)_C\times U(2)_L \times U(1)_R \times
U(1)_N$ arises from breaking an unified group $G=U(7)$
whose adjoint have the following charge assignment under the above
subgroup $H$,
\begin{equation} \label{su7adj} {\bf 48} =
\begin{pmatrix}
\bf 8 & q & u & \\
q^{\rm c}   & \bf 3 &   & l \\
u^{\rm c}   &   &  \bf 1 & e \\
   & l^{\rm c}  & e^{\rm c}  & \bf 1 \\
\end{pmatrix}
\end{equation}
where the notations are self-explanatory; the block-diagonal
numbers refer to the dimensions of gauge bosons, and off-diagonal
blocks correspond to complex squarks and sleptons and
hermitian conjugates. We will see the fields corresponding to blank entries are
massive. They have the correct quantum
numbers, as well as hypercharge defined as linear combinations
\begin{equation} \label{ycharge}
 Q_Y = \frac16 Q_C - \frac12  Q_L -\frac12 Q_R.
\end{equation}
This breaking is achieved by the following
background magnetic flux $F = 2 \partial_{[4}A_{5]}$ on the extra
two-torus $T^2$,
\begin{equation} \label{toyBG} \begin{split}
 2 \pi \alpha' A_5 =  &
 \begin{pmatrix} {m_1 \over n_1} I_{n_1} & & &  \\ &
 {m_2 \over n_2} I_{n_2} & & \\ & &  {m_3 \over n_3} I_{n_3} & \\ & &
 &  {m_4 \over n_4} I_{n_4}
 \end{pmatrix} x^4 \\
& +
 \begin{pmatrix} a_1 I_{n_1} & & &  \\ &  a_2 I_{n_2} & & \\ & &
 a_3 I_{n_3} & \\ & &  &  a_4 I_{n_4}
 \end{pmatrix}
.
\end{split}
\end{equation}
Here gcd$(n_a,m_a)$ are $3,2,1,1$, respectively,
$m_1/n_1=m_3/n_3,m_2/n_2=m_4/n_4,a_1 \ne a_3,a_2 \ne a_4$, and
$\alpha'$ will be interpreted as Regge slope. In compactification with
more dimensions, we may put some of the blocks to different gauge field
components than $A_5$ \cite{Ra}.

Most of all, the theory is {\em chiral} \cite{CIM,AW}. This
means, for example either quark $q$ or antiquark $q^{\rm c}$ in
(\ref{su7adj}) will be exclusively massless. This can be seen by index theorem.
The Dirac operator is decomposed as $\Ds_6 = \Ds_4 +
\Ds_2$ and, for $4k+2$ extra dimensions, the chirality is correlated.
The difference of the number of left and right mover zero modes, is
the first Chern number,
\begin{equation} \label{diracidx}
n_+ - n_- = {1 \over 2 \pi} \int_{T^2} \Tr_Q F,
\end{equation}
where the trace is over gauge charge $Q$ of commutant group $L$ to
$H$, completely determined by branching. Plugging
(\ref{toyBG}), we see that the each quark and lepton is chiral.
Moreover, there are $n_+ - n_-$ degenerate
bifundamental solutions for each off-diagonal solution. One can confirm
that this is the same as intersection number $m_a n_b - n_a m_b$ of D-branes,
which can explain the number of generations. It is determined by the
topology of extra dimensions. Note that gauge bosons of $H$ are always
massless, nonchiral and non-degenerate.

Such a constant magnetic flux $F$ is understood, in the
$T$-dual space, as linearly growing vacuum expectation values (VEVs)
of adjoint scalar field $X^{5 \prime}$ up to a gauge choice,
describing tilted brane with
respect to coordinate axis. The boundary condition
of torus requires $m_a, n_a \in {\bf Z}$ \cite{tH}. These numbers $(n_a,m_a)$
correspond to winding numbers (homology cycles) of D-branes.

These patterns of symmetry breaking are achieved via two kinds of
mechanism, in the brane language:
\begin{itemize}
\item parallel separation of stacks: the constant part
\item partial brane stack ``rotation'': the linear part
\end{itemize}
cf. Eq (\ref{toyBG}).
The first is what we usually have in Higgs mechanism with a constant
adjoint VEV $\langle A_0 \rangle$ \cite{BLS}: All the fields
corresponding to off-diagonal entries get massive. The
second is described by developing constant {\em flux} on gauge field strength
\begin{equation} \label{growvev}
 \langle A \rangle  = \langle F \rangle x + \langle A_0 \rangle.
\end{equation}
The resulting unbroken symmetry is the one commutes with (\ref{growvev}).
By the mechanism discussed, only the fermions with one chirality
survive, which is determined by the orientation of flux or brane rotation.
In fact the continuous rotation is not possible, because the quantization
condition of $(n_a,m_a)$. By ``rotation'' we mean recombination which
we will discuss shortly.

The example (\ref{su7adj}) thus is obtained as follows, where $T$-dual
picture is more transparent. Start with 7 slices of coincident branes,
separate stacks with $3_C+2_L+1_R+1_N$ branes. Then after rotating
$2_L$ and $1_N$ branes, we obtain
chiral spectrum. One can check \cite{GR} that the commuting generators
of $U(7)$ represents unbroken gauge group and chiral quarks and
leptons, exclusively not paired with the charge
conjugate, indicated in (\ref{su7adj}).

We will generalize this argument to $4k+2$ extra dimensions, among
them our interest is six. We have seen that it is easy to obtain
chiral fermion zero mode, but only with supersymmetry the massless boson exists
as superpartner. The supersymmetry condition is moduli dependent.
In fact, for the case of 2 extra dimensions, there can
never be massless scalar. In what follows, we will consider higher
dimensional theory and rotation {\em preserving supersymmetry.}

The above toy model is anomalous. The best explanation for an anomaly
free theory is that it is spontaneously broken phase of a unified
theory, where the absence of anomaly is natural. The most suggestive
scenario will be ten dimensional supergravity with nonabelian
gauge group $SO(32)$, which is the only anomaly free theory
if we want the perturbative open string description. Symmetries are broken by
projections associated with the compactification. In the brane picture,
consistency is given by Ramond--Ramond (RR) tadpole cancellation condition.
It guarantees anomaly freedom of nonabelian gauge anomalies and
of other mixed anomalies involving $U(1)$'s by
the generalized Green--Schwarz (GS) mechanism from
antisymmetric tensor fields.

For this we introduce an orientifold plane (O-plane) having opposite
RR charges to D-branes. This will set an upper bound of the rank of
the gauge group. This is
obtained by compounding worldsheet parity reversal
$\Omega$ and spatial reflection, under which O-plane is fixed.
This brings about mirror fermion having charge $\bf (M,N)$ for $\bf
(M,\overline N)$ corresponding to mirror cycles. When a brane and its
image brane are on top of orientifold plane, the gauge group is $SO$ type.
The symmetry breaking $SO(A)\times U(B) \subset SO(A+2B)$ is again described
by parallel separation.
The $2\times 2$ block matrix $\left( \begin{smallmatrix} 0
  & 1 \\ -1 & 0 \end{smallmatrix} \right)$ plays the role of imaginary
number, determined by Chan--Paton projector. For
example, the above $SU(7)$ toy model is realized and embedded in
$SO(14) \subset SO(32)$, having almost same structure. Since the final
group is $SO(32)$, inevitably we have hidden sector \cite{Ni}.

During the compactification, we may meet more than one gauge groups
thus as many adjoints, depending on the number of O-planes, so all the
quarks and leptons are not necessarily belong to the same adjoint.

\section{Brane recombination}

The above gauge symmetry breaking is studied as deformations of
branes, ``brane recombinations''. Mainly it has been
discussed in non-supersymmetric theory in the context of tachyon condensation.
However we want supersymmetry preserving recombination.
In the brane picture, the supersymmetry exists when the cycles belong
to $SU(3)$ holonomy special Lagrangian. The
supersymmetry-preserving deformation is parameterized by Wilson lines,
as a special case of McLean's theorem \cite{Mc,GR}. In particular, turning on
off-diagonal components changes the cycle and ``rotates'' the brane.
In other words, although we cannot continuously change the linear
entries in (\ref{toyBG}) due to integral quantization condition of
$m_a,n_a$, we can change them ``effectively'' \cite{EGHK,HT,CIM02}.
It is because the corresponding fields to off-diagonal components are
charged under the subgroup $L$, which breaks $G$ down to the commutant $H$.
In this sense, the description as (\ref{toyBG}) is {\em far from
complete}.
The proper description comes with coherent sheaves in K-theory
\cite{MM}, and we just use it as the resulting ``effective'' phase of
broken theory.

In any recombinations, in fact in any interactions of D-branes,
RR charges should be conserved.
It means every recombination takes place preserving the {\em anomaly
  content.} When we survey $F \wedge F$ term of Chern class as in
(\ref{diracidx}), it is invariant under brane recombinations.
Since we are considering
$4k+2$ dimensions, an adjoint (before recombination), as well as chiral
fermions, contributes chiral anomaly. After symmetry breaking, this
number is invariant. As long as tadpole is cancelled, there is no
chiral anomalies in the low energy limit.

\begin{figure}
\includegraphics[height=2cm]{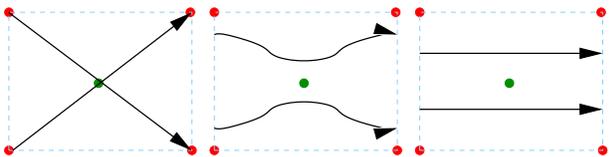}
\caption{Brane recombination (only one two-torus is shown).
  Every intermediate setup satisfies the
  same 1/4 BPS condition (\ref{qtBPS}) thus there is no energy cost
  on deformation.}
\end{figure}
To be specific, let us consider D6 branes in
type IIA theory. We consider a ``1/4''(BPS)-cycle, where only two
cycle is wrapping nontrivially on four torus $T^4$ in, say, 4567 direction
and the remaining one cycle is along the O-plane in 89
direction. Now consider a recombination \cite{Chtype1}
\begin{equation*} \begin{split}
&(n_1,m_1)(n_2,m_2)(1,0)+(n_1,-m_1)(n_2,-n_2)(1,0) \\
 \to & 2n_1 n_2(1,0)(1,0)(1,0)+2m_1 m_2(0,1)(0,1)(1,0).
\end{split} \end{equation*}
In the presence of O-planes, we can choose two cycles as
being {\em always the images with respect to an O-plane}, here
(1,0)(1,0)(1,0), and this reflection
symmetry is always preserved. Supersymmetry (seen by worldsheet spectum)
requires the intersecting angles to be same $\theta_2=\theta_3$,
which translates to
$$F_{45} = F_{67} \propto \left(
  \begin{matrix} {m_1 \over n_1}I_{n_1 n_2} &  \\  & -{m_1 \over
      n_1}I_{n_1 n_2} \end{matrix}
 \right), F_{89}=0,$$
which the above cycles satisfy.
We observe the preserved supersymmetry component $\epsilon$
\begin{equation} \label{qtBPS}
\Gamma^{\mu \nu} F_{\mu \nu} \epsilon = 0
\end{equation}
is same for every intermediate step. It is because this relation is local, so
that at every point of intermediate brane state this condition is
satisfied.
Since supersymmetry is preserved, there is no energy cost on ``marginal''
deformation. We can verify it by explicit DBI energy relation from
the above supersymmetry condition
\begin{equation} \label{energies} \begin{split}
   & \Tr  \tau_9  V_{2} V_3 V_4 \sqrt{(1+ (2 \pi \alpha' F_{45})^2)(1+
   (2 \pi \alpha' F_{67})^2)} \\
 & = n_1 n_2  \tau_9 V_2 V_3 V_4 +  m_1 m_2 \tau_5 V_4
\end{split} \end{equation}
with $\tau_p \propto (2\pi \sqrt{\alpha'})^{-p}$ being D$p$ brane tension and $V_i$ volume of $i$th two-torus.
In $T$-dual picture in $x^5,x^7,x^9$ directions, the second line shows
that the setup, before and after recombination,
has exactly the same energy as D5 branes on top of D9 branes,
aligned on top of O-planes, which is again in the original picture:
all the intersecting D6s {\em along the O-planes}. Therefore the energy
of branes, before and
after recombinations, are {\em same.}
The interpolating dynamics is $T$-dual to D0-D4 bound state, or
instanton obeying peroidicity ``toron.'' The deformation freedom in
this example corresponds to the two $SU(n_1 n_2)$ instanton moduli
space \cite{AH,GR}.

Although we have seen that there is a recombination
process that deforms intersecting branes to parallel ones, gauge
coupling changes during this process.
The four dimensional gauge coupling is proportional to  $V\sqrt {-\det
  {(1+2 \pi \alpha' F)_\mu}^\nu}$ for each stack, with a background
flux $F$ provided by $T$-dual picture of branes, which can be seen by expanding
fluctuation around these flux $A_\mu = \langle A_\mu \rangle + \delta
A_\mu$ and reading off YM coupling by canonical noramlization
\cite{AHT}. This is an unique property of DBI action which is
not present in YM description.
Thus below the compactification scale the
intersecting branes does not give unified coupling, in general.
This does not conflicts with neither energy
costless deformation since it is the dynamics of extra dimensions, nor
unification picture above unification scale. In
the four dimensional effective theory, running from the low energy,
the gauge coupling unification is modified by large threshold
correction around unfication scale. The
backgroun value $F^2 \sim {\cal O}(\alpha^{\prime-1}) $ above is order
of compactification scale. This means the fluctuation is
not small any more compared to the background, and above unification
scale opens up extra dimensions with the usual unified gauge coupling
from type I string.

The above D0-D4 bound state is $U$-dual to (F,D$p$) bound state
which is interpreted as the electric flux on D-branes
\cite{HT,CK}. The above BPS equation becomes string junction
condition \cite{DM}
and we can clearly see the supersymmetry condition at every local point, with
the same supersymmetry components preserved.
In this picture we can also see the no-energy-cost property under marginal
deformation and the specific shape of final state \cite{CK}. The
DBI energy is $V\sqrt {-\det {(1+2 \pi \alpha' F)_\mu}^\nu}.$
When we have purely magnetic flux, this is nothing but the
winding volume of torus $V\sqrt{1+(2\pi \alpha' F)^2} =
V\sqrt{1+\tan^2 \theta} = V_{\rm cycle}$, agreeing with intersecting brane
case. But when we have contribution from fundamental string, which
carries NS-NS flux, which is equivalent to electric field component.
The size of electric and magnetic fluxes should be equal due to BPS
condition, so the determinant becomes 1, which shows independence of
cycle.

Another description is from the algebraic geometry \cite{DZ}.
Such special Lagrangian configuration is described by
algebraic variety and deformation is described by interpolating
curve. All the above 1/4-cycles can be embedded into complex curves,
which are always special Lagrangian cycles in two complex dimension.

For less symmetric cycles such as 1/8 cycles, there still be the
recombination process, which is also marginal, protected by
supersymmetry \cite{BBH}. However in this case, the parallel brane
state has more supersymmetry, namely 1/4 BPS. We recall that in the
nonsupersymmetric case, a runaway minimum has a recovered supersymmetry.
It is clear that the conservation of RR charge should drive the state
again to that with parallel branes all on top of orientifold planes.

When more than one adjoint Higgs assume VEVs, in general the rank of
the gauge group
reduces. An important application is the electroweak Higgs mechanism
\cite{CIM02}. It occurs when the bifundamental Higgs assumes constant
VEV, which corresponds to assigning off-diagonal VEVs to original adjoint
representation (\ref{su7adj}), in the position of $l$.

\section{Type I compactifications}

Finally we argue that all the SUSY type I/II orientifold models
constructed so far be continuously connected, by above deformations and
$T$-duality, {\em to type I theory compactified on orbifold}.
The vital constraint is RR tadpole cancellation condition.

To cancel tadpole, it is crucial that every
construction requires {\em at least one} orientifold plane O$p$ as a
negatively RR charged object. This can always be converted to O9 plane by
$T$-duality. In general, an orientifold group is generated by orbifold
actions $g \in P$ and worldsheet parity reversals $\Omega \prod R_m
(-1)^{F_L}$,
where $R_m$ are spatial reflections of compact $x^m$ directions and
$F_L$ is the spacetime fermion number. This O$p$
plane is lying on the fixed plane under $\prod R_m$.
By the $T$-dual $\prod R_m (-1)^{F_L}$, we can always convert it to $\Omega$,
while orbifold group $P$ is untouched.
The type IIB theory with orientifold element $\Omega$ is
type I theory, so that every theory containing orientifold is regarded as
compactification of type I string theory with $P$.
It follows that the {\em orbifold group $P$}
completely determines additional orientifolds and thus gauge group.

Fixing one orientifold as O9, let us then count the lower dimensional
orientifolds. Consider a $T^n/\Z_N$ orbifold \cite{AFIV}. For odd $N$, the
only possible orientifold is O9 since there is no even order element
in $P$ compatible with lower dimensional orientifold. For even $N$, the only
possibility other than O9 is O5, because in order
not to have a tachyon, the difference of dimensions of D-branes
should be a multiple of 4. We cannot put an additional O1, if we want four
dimensional Lorentz invariance. The number of O5 is always 1, and it sits on
the plane generated by order 2 element $g^{N/2}$. Note that if this
orientifold group element $\Omega g^{N/2}$ exists, introducing O5 is
compulsory, which would be absent in the untwisted theory. The same is
also applied to $T^n/(\Z_M \times \Z_N)$
orbifolds, where the number of O5 can be up to 3.
Other systems containing O$p$,D$p$ and/or O$(p-4)$,D$(p-4)$ can be
$T$-dualized to the above models.

One can see, for example, the model in \cite{CSU} is deformable to the one
in \cite{BL}, both of which are based on $T^6/(\Z_2\times \Z_2)$
orbifold. In
the latter picture, orbifold actions determine three $5+1$ dimensional
hyperplanes and harbor O-planes. With Chan--Paton projectors, the
resulting gauge group consists of $Sp(k)$, which
is embedded in $SO(32)$, for each O5 plane. $SO(32)$ because
the product of the RR charge of O$p$ and the number of fixed points is
always 16. In the maximal group case
(without Wilson lines, or when all the D-branes are on top of one
orientifold planes) we have only gauge group $Sp(8)^4$ which will be the
final unficiation group obtainable from brane
deformations. Nevertheless they are to be regarded as {\em broken $SO(32)^4$} because
the resulting 99 and $5_i 5_i$ spectra in \cite{BL}
can be explained by branching rules ${\bf 496 \to 136} + 3 \cdot {\bf
 120}$. Note also we have
$T$-dual symmetries exchanging O9 with any of the O5$_i$.
Therefore every gauge group arising from D$p$-D$p$
branes can be {\em embedded into  $SO(32)$, for each orientifold plane}.
As a bonus, we see the $Sp(k)$ group describing small instantons
\cite{Wi95} is embedded into $SO(32)$, too.

Applying $T$-dualities in $x^5,x^7,x^9$ directions we have four O6s.
By brane deformations, we can see the spectrum of \cite{CSU} can be
obtained from this.
We can always find the directions in which $T$-duality maps all
O9,D9,O5,D5 to six dimensional objects: O6 and D6.
Therefore we can lift of the model to $M$-theory on $G_2$ manifold
picture, where the desirable objects are six dimensional objects
in Type IIA theory, which become geometric objects, i.e. singularities.

One may note that not every vacuum might be connected, since the
deformation of orbifold/orientifold images should be always deformed
together. However it is strongly restricted by anomaly cancelation of
representations of $Spin(32)/\Z_2$ \cite{BLPSSW}.

\acknowledgements

The author is grateful to Hans-Peter Nilles, Piljin Yi, Chengang Zhou
and especially to Daniel Cremades for discussions and correspondences.
This work was partially supported by the European Union 6th Framework
Program MRTN-CT-2004-503369 Quest for Unification and
MRTN-CT-2004-005104 ForcesUniverse.

\end{document}